\documentclass{emulateapj}

\pdfoutput=1
\usepackage{hyperref}
\usepackage{graphicx}
\usepackage{amssymb}

\begin{document}
\submitted{August 14, 2008}
\title{Measurement of the dark matter velocity anisotropy in galaxy clusters}
\author{Ole Host\altaffilmark{1}, Steen H.~Hansen\altaffilmark{1}, Rocco Piffaretti\altaffilmark{2}, Andrea Morandi\altaffilmark{1,3},  Stefano Ettori\altaffilmark{4}, Scott T.~Kay\altaffilmark{5}, Riccardo Valdarnini\altaffilmark{2}}
\altaffiltext{1}{Dark Cosmology Centre, Niels Bohr Institute, University of Copenhagen, Juliane Maries Vej 30, DK-2100 Copenhagen, Denmark}
\altaffiltext{2}{SISSA, International School for Advanced Studies, via Beirut 2, I-34014 Trieste, Italy}
\altaffiltext{3}{Dipartimento di Astronomia, Unversit\`a di Bologna, via Ranzani 1, I-40127 Bologna, Italy}
\altaffiltext{4}{INAF--Osservatorio Astronomico di Bologna, via Ranzani 1, I-40127 Bologna, Italy}
\altaffiltext{5}{Jodrell Bank Centre for Astrophysics, School of Physics and Astronomy, The University of Manchester, Manchester M13 9PL, UK}

\begin{abstract}
The internal dynamics of a dark matter structure may have the remarkable property that the local temperature in the structure depends on direction. This is parametrized by the velocity anisotropy $\beta$ which must be zero for relaxed collisional structures, but has been shown to be non-zero in numerical simulations of dark matter structures. Here we present a method to infer the radial profile of the velocity anisotropy of the dark matter halo in a galaxy cluster from X-ray observables of the intracluster gas. This non-parametric method is based on a universal relation between the dark matter temperature and the gas temperature which is confirmed through numerical simulations. We apply this method to observational data and we find that $\beta$ is significantly different from zero at intermediate radii. Thus we find a strong indication that dark matter is effectively collisionless on the dynamical time-scale of clusters, which implies an upper limit on the self-interaction cross-section per unit mass $\sigma/m\lesssim1\,$cm$^2$g$^{-1}$. Our results may provide an independent way to determine the stellar mass density in the central regions of a relaxed cluster, as well as a test of whether a cluster is in fact relaxed.\end{abstract}

\section{Introduction}

Our understanding of dark matter structures has increased significantly over the recent years. This progress has mainly been driven by numerical simulations which have
identified a range of universalities of the dark matter structures. One of the first general
properties to be suggested was the radial density profile \citep{1996ApJ...462..563N,1998ApJ...499L...5M,2004MNRAS.353..624D,2006AJ....132.2685M,2006AJ....132.2701G}, whose radial behaviour was shown to change from a fairly shallow decline in the central region to a much steeper decline in the outer regions. This behaviour has been confirmed observationally for galaxy clusters, both through X-ray observations \citep{2004ApJ...604..116B,2005A&A...435....1P,2005A&A...441..893A,2006ApJ...640..691V,2006A&A...446..429P}, and also more recently through strong and weak lensing observations \citep{2004ApJ...604...88S,2005ApJ...619L.143B,2006ApJ...642...39C,2008arXiv0802.4292L}.

A slightly less intuitive quantity to be considered is the dark matter velocity anisotropy defined by
\begin{equation}\label{eq:be}
\beta \equiv 1 - \frac{\sigma^2_t}{\sigma^2_r},
\end{equation}
where $\sigma^2_t$ and $\sigma^2_r$ are the 1-dimensional tangential and radial velocity dispersions in a spherical system~\citep{1987gady.book.....B}. This anisotropy was shown in pure dark matter simulations to increase radially from zero in the central region to roughly 0.5 in the outer region~\citep{1997ApJ...485L..13C,1996MNRAS.281..716C,2006NewA...11..333H}. For collisional systems, in contrast, the velocity anisotropy is explicitly zero in the equilibrated regions. Therefore, eventually inferring $\beta$ from observational data is an important test of whether dark matter is in fact collisionless, as assumed in the standard model of structure formation. On this note, it has been shown that the Galactic velocity anisotropy can affect the detection rates of direct dark matter searches \citep{2008PhRvD..77b3509V}, and it is in principle measurable in a direction-sensitive detector \citep{2007JCAP...06..016H}.

The most massive bound structures in the Universe are clusters of galaxies, which consist of an extended dark matter halo, an X-ray emitting plasma making up the intracluster medium (ICM), and the individual galaxies. While the contribution of galaxies to the total mass is small, approximately 10 $\%$ of the cluster mass resides in the ICM. The present generation of X-ray satellites, \emph{XMM-Newton} and \emph{Chandra}, allows very accurate measurements of azimuthally-averaged radial profiles of density and temperature of the ICM. These are used, under the assumption of hydrostatic equilibrium and spherical symmetry of both gas and total mass distributions, to estimate total, gas, and dark matter mass profiles \citep{1980ApJ...241..552F}.

Below we infer the radial velocity anisotropy profile of dark matter in 16 galaxy clusters using a generally applicable framework without any parametrized modeling of the clusters. In short, we assume a universal relation between the effective temperature of dark matter and the ICM temperature, which allows us to solve the dynamics of the dark matter halo using the radial gas temperature and density profiles determined from X-ray data. We investigate the shape and validity of this temperature relation in two cosmological simulations of galaxy clusters, based on independent numerical codes. We apply our method to 16 galaxy clusters from two different samples and find a velocity anisotropy significantly different from zero in the outer parts, in qualitative agreement with simulations. 

Our approach here is a generalization of the non-parametric analysis in \citet{2007A&A...476L..37H} where $\beta$ was inferred neglecting the radial dependence. We also note the parametrized analyses in \citet{2004ApJ...611..175I} and \citet{2007MNRAS.380.1521M}, where the total dark matter velocity dispersion was inferred assuming either $\beta=0$, or the analytical $\beta$-profiles of \citet{2000ApJ...539..561C} or \citet{1996MNRAS.281..716C} (see also \citet{2008MNRAS.tmp..719W}). In particular, \citet{2007MNRAS.380.1521M} found that the dark matter temperature and the ICM temperature were essentially the same in strong cooling-core clusters.

The structure of the paper is the following: In the next section, we discuss how we relate the temperature of dark matter to the observable gas temperature. In section 3 we show how we can then solve the dynamics of the dark matter. In section 4 we test the assumed temperature relation and our method on numerical simulations, and in section 5 we apply the method to observational data. Section 6 is the summary and discussion.

\section{The temperature of dark matter}
The equality of inertial and gravitational mass implies that the orbit of a test particle in a gravitational system is independent of mass. For example, the velocity of a circular orbit in a spherical mass distribution $v_c^2=GM(r)/r$ depends only on the distance to the center of the system and the mass contained within that radius. Therefore it is natural to assume that, at a given radius, all species in a relaxed, spherical gravitational system have the same average specific kinetic energy. Obviously, they also have the same specific potential energy. In a gas system, equilibrium implies energy equipartition between all species. It is clear that the corresponding principle for a relaxed gravitational system is a common velocity dispersion, precisely because gravitational dynamics are independent of mass. Since the average velocity is associated with the thermal energy content, this relationship is expressed by
\begin{equation}\label{eq:trl}
T_{\mathrm{DM}}=\kappa T_\mathrm{gas}.
\end{equation}
The parameter $\kappa$ is constant as long as the impact of radiative or entropy-changing processes affecting the gas is negligible and the system is relaxed. Therefore, we allow for a radial dependence, $\kappa=\kappa (r/r_{2500})$, where $r_{2500}$ is the scale radius within which the mean total density is $2500$ times the critical density at the redshift of the cluster. 

The dark matter temperature in (\ref{eq:trl}) is naturally not well-defined as there is no thermodynamic equilibrium for a collisionless gas. Instead, we simply define an effective dark matter temperature which is proportional to the three-dimensional velocity dispersion,
\begin{eqnarray}
k_BT_{\mathrm{DM}}&=&\frac{1}{3}\mu m_H\sigma^2_\mathrm{DM}\\
&=&\frac{1}{3}\mu m_H\left(\sigma_r^2+2\sigma_t^2\right).\label{eq:tdm}
\end{eqnarray}
The velocity dispersion has been decomposed into the contributions from the one-dimensional radial and tangential dispersions. We choose the constant of proportionality to be the mean molecular mass of the intracluster gas simply to allow $\kappa$ to be of order unity. Equations (\ref{eq:trl})--(\ref{eq:tdm}) are equivalent to assuming that the specific energies of gas and dark matter particles are the same up to a factor of $\kappa$, on average. The same conjecture was made in \citet{2007A&A...476L..37H} but with $\kappa=1$ explicitly. 

It should be mentioned that the temperature relation (\ref{eq:trl}) was recently analyzed in simulations by \citet{2008ApJ...672..122E}. Whereas we allow a possible radial variation in the temperature relation, those authors considered averages within $r_{200}$ and found that
\begin{equation}
\tilde{\kappa}_{<r_{200}}\equiv\frac{k_B\overline{T_\mathrm{gas}}/\mu m_H}{\overline{\sigma^2_\mathrm{DM}}}=1.04\pm0.06,
\end{equation}
This was based on their determination of $\bar{\kappa}^{-1}_{<r_{200}}=(0.87\pm0.04)\langle T_\mathrm{spec}/T_\mathrm{mw}\rangle$, where the ratio of the spectroscopic temperature to the mass-weighted temperature was $\langle T_\mathrm{spec}/T_\mathrm{mw}\rangle=1.1\pm0.05$ \citep{2006ApJ...650..538N}. Then, in \citet{2008ApJ...679L...1R} it was noted that by applying virial scaling to the WMAP5+SN+BAO results \citep{2008arXiv0803.0547K}, an average value of $\tilde{\kappa}^{-1}|_{<r_{500}}=1.1$ was found. The authors concluded that the observational results indicated that the average specific energy of the ICM was close to both that of the dark matter and that of the galaxies. In section \ref{sec:sim}, we will arrive at the same conclusion for simulated galaxy clusters. 

\section{Solving the dark matter dynamics}\label{se:th}
Equation (\ref{eq:trl}) allows us to estimate the total velocity dispersion profile of the dark matter structure from measurements of the radial temperature profile of the gas. In this section we discuss how we can proceed to determine the dark matter velocity anisotropy.

The collisionless Jeans equations relate the dynamical properties of the dark matter to the gravitational potential of the cluster. Assuming that the system is spherically symmetric and in a steady state, the second Jeans equation can be put in the form \citep{1987gady.book.....B}
\begin{equation}
\frac{d (\nu\overline{v_r^2})}{dr}+\frac{\nu}{r}\left[2\overline{v_r^2}-\left(\overline{v_\theta^2}+\overline{v_\phi^2}\right)\right]=-\nu\frac{GM}{r^2},
\end{equation}
where $\nu$ is the dark matter number density, $\overline{v_i^2}$ is the second moment of the $i$th velocity component, and $M$ is the mass contained within radius $r$. If it is further assumed that there are no bulk flows, $\overline{v_i}=0$, and that the tangential velocity dispersions are equal, $\sigma_\theta^2=\sigma_\phi^2\equiv\sigma_t^2$, the Jeans equation becomes
\begin{equation}\label{eq:je}
\sigma_r^2\left(\frac{d\ln \rho_{\mathrm{DM}}}{d\ln r}+\frac{d\ln \sigma_r^2}{d\ln r}+2\beta\right)=-\frac{GM(r)}{r},
\end{equation}
where $\rho_\mathrm{DM}$ is the mass density, $\sigma_r^2$ is the radial velocity dispersion, and $\beta$ is the velocity anisotropy introduced in (\ref{eq:be}).

Similar to the Jeans equation, the radial part of the Euler equations of the ICM expresses the condition that the thermal pressure of the gas balances the gravitational potential. This equation of hydrostatic equilibrium reads
\begin{equation}\label{eq:he}
\frac{k_BT_\mathrm{gas}}{\mu m_H}\left(\frac{d\ln n_e}{d\ln r}+\frac{d\ln T_\mathrm{gas}}{d\ln r}\right)=-\frac{GM(r)}{r},
\end{equation}
where $T_\mathrm{gas}$ is the gas electron temperature and $n_e$ is the number density of electrons. This important equation has been widely used to estimate the total mass of a galaxy cluster from X-ray data. In case there is turbulence or larger scale bulk motion in the gas  additional terms of the form $(\vec{v}\cdot\nabla)\vec{v}-(v_\theta^2+v_\phi^2)/r$ appear \citep{lanlif}. Neglecting such terms may lead to an underestimate of the mass, however this is usually not a major effect for systems that appear relaxed \citep{2008arXiv0808.1111P}.

By equating (\ref{eq:je}) and (\ref{eq:he}) and using (\ref{eq:be}) and (\ref{eq:trl}) to eliminate $\beta$, we obtain the following differential equation for the radial velocity dispersion,
\begin{equation}\label{eq:rd}
\sigma_r^2\left(\frac{d\ln \rho_{\mathrm{DM}}}{d\ln r}+\frac{d\ln \sigma_r^2}{d\ln r}+3\right)=\psi(r),
\end{equation}
where the function $\psi$ is defined by
\begin{equation}
\psi(r)=3\kappa\frac{k_BT_\mathrm{gas}}{\mu m_H}-\frac{GM}{r}.
\end{equation}
Clearly, $\psi$ is determined directly from the X-ray observables and the $\kappa$-profile, which we discuss in section \ref{sec:simtrl}.

The differential equation (\ref{eq:rd}) is solved by finding an integrating factor which yields
\begin{equation}\label{eq:srsq}
\sigma_r^2(r)=\frac{1}{\rho_{\mathrm{DM}}(r)\,r^3}\int_0^r dr'\, \psi(r')\rho_{\mathrm{DM}}(r')r'^2.
\end{equation}
The dark matter density is determined as usual through $\rho_{\mathrm{DM}}=\rho_{\mathrm{tot}}-\mu m_H n_e$. With the radial velocity dispersion profile determined, the velocity anisotropy is easily recovered from either the temperature relation (\ref{eq:trl}) or the Jeans equation (\ref{eq:je}),
\begin{eqnarray}
2\beta_{\mathrm{tr}}&=&3\left(1-\kappa\frac{k_BT_\mathrm{gas}}{\mu m_H \sigma_r^2}\right),\label{eq:btr}\\
2\beta_{\mathrm{Je}}&=&-\frac{d\ln(\rho_{\mathrm{DM}}\sigma_r^2)}{d\ln r}-\frac{GM}{r\sigma_r^2}.\label{eq:bje}
\end{eqnarray}
Obviously these two expressions should be equal. This can be used as a consistency check on whether numerical issues related to the differentiations and integration involved are kept under control. 

To summarize, the assumed relation (\ref{eq:trl}) between the effective dark matter temperature and the gas temperature, along with the mass estimate from (\ref{eq:he}), allows us to solve the dark matter dynamics directly from X-ray data, and determine both the radial velocity dispersion and the velocity anisotropy as functions of radius. 

\section{Cluster simulations}\label{sec:sim}
We use numerical simulations of the formation of galaxy clusters in the $\Lambda$CDM cosmology to investigate the validity and shape of the temperature relation (\ref{eq:trl}), and to test the method for determining the velocity anisotropy. In order to check systematic effects we take samples from two different simulations based on two completely independent numerical codes.

\subsection{CLEF}
We first consider a sample of 67 clusters taken from the CLEF simulation \citep{Kay:2006iz}, details of which are briefly summarized here. The CLEF simulation was run with the GADGET2 $N$-body/SPH code \citep{2005MNRAS.364.1105S} and followed the evolution of large-scale structure within a box of comoving length, $200 h^{-1}{\rm Mpc}$. The following cosmological parameters were assumed: $\Omega_{\rm m}=0.3; \Omega_{\Lambda}=0.7; \Omega_{\rm b}=0.0486; h=0.7; n_{\rm s}=1; \sigma_8=0.9$. Here the value of the Hubble constant is written as $100 \, h\,\mathrm{km}\,\mathrm{s}^{-1}\,\mathrm{Mpc}^{-1}$ and $\sigma_8$ is the rms mass fluctuation at the present epoch in a sphere of radius $8\,h^{-1} \mathrm{Mpc}$. The number of particles was set to $428^3$ for each of the gas and dark matter species, thus determining the particle masses to be $m_{\rm DM}=7.1\times 10^{9}h^{-1}{\rm M}_{\odot}$ and $m_{\rm gas}=1.4\times 10^{9}h^{-1}{\rm M}_{\odot}$ respectively. The equivalent Plummer softening length was set to $20\,h^{-1}\,$kpc and held fixed at all times in comoving co-ordinates. Pressure forces were calculated using the standard GADGET2 entropy-conserving version of SPH with an artificial viscosity to convert kinetic energy into thermal energy where the flow was convergent. The gas could cool radiatively assuming a fixed metallicity, $Z=0.3Z_{\odot}$. Cold ($T<10^{5}$K) gas with $n_{\rm H}>10^{-3}{\rm cm}^{-3}$ either formed stars or was heated by an entropy, $\Delta S=1000\, {\rm keV \, cm}^2$. This choice was determined stochastically by selecting a random number, $r$, from the unit interval and heating the particle if $r<0.1$, i.e. a 10 per cent probability of being heated. This high level of feedback was necessary to reproduce the observed excess entropy in clusters (see \citet{Kay:2006iz} for further details).

To select the cluster sample, we first consider all clusters at $z=0$ with X-ray temperatures, $kT>2\, {\rm keV}$; this produces 95 objects, with virial masses, 
$M_{\rm vir}>1.3\times 10^{14}h^{-1}{\rm M}_{\odot}$ (correspondingly, $> 15,000$ dark matter particles). We then select those clusters with 3D substructure statistic, $s<0.05$. The substructure statistic \citep{1998MNRAS.296.1061T} measures the displacement of the centre of mass from the potential minimum of the cluster (taken to be its centre), relative to $r_{500}$, which is the scale radius within which the mean total density is $500$ times the critical density. By making this cut, we therefore exclude all clusters that show significant signs of dynamical activity, i.e. major mergers. 

\subsection{V06}
The second sample is a subsample of the one presented in \citet{2006NewA...12...71V} which we refer to as V06. These simulations assumed a concordance flat $\Lambda$CDM with the same cosmological parameters as for the CLEF simulation.

The simulation ensemble of galaxy clusters was constructed according to
a procedure described in \citep{2008arXiv0808.1111P}. Here we briefly summarize the most important aspects. The hydrodynamic simulations were run using an entropy-conserving multistep TREESPH code for a sample of 153 clusters  spanning a range from $\simeq 1.5 \times 10^{15} h^{-1} M_{\odot}$ down to $M_{vir}\simeq 1.5 \times 10^{14} h^{-1} M_{\odot}$. The initial conditions ($z_{in}=49$) were extracted from a set of purely N-body cosmological simulations in which clusters of galaxies were identified from the particle distribution at $z=0$ using a friends--of--friends algorithm. In order to investigate the effect of the implemented gas processes on the energy equipartition between gas end dark matter particles, we performed both adiabatic and radiative simulations. The radiative simulations are of course more realistic than the adiabatic ones, because they additionally take into account radiative cooling, star formation, energy and metal feedback \citep{2003MNRAS.339.1117V}. More details concerning the simulation technique and the implementation of physical processes of the gas are given in \citet{2006NewA...12...71V}.

In order to avoid contamination from dynamically perturbed clusters, we select the 20 most relaxed objects at $z=0$. The selection is based on the power ratio method, which measures the amount of substructure in X-ray surface brightness maps. The map sources a pseudo-potential which is expanded in plane harmonics, and the ratio of the third coefficient to the zeroth is a measure of substructure. More details are given in \citet{2008arXiv0808.1111P}.  

\subsection{The temperature relation}\label{sec:simtrl}

\begin{figure}[tbp]
\begin{center}
\epsscale{1}
\plotone{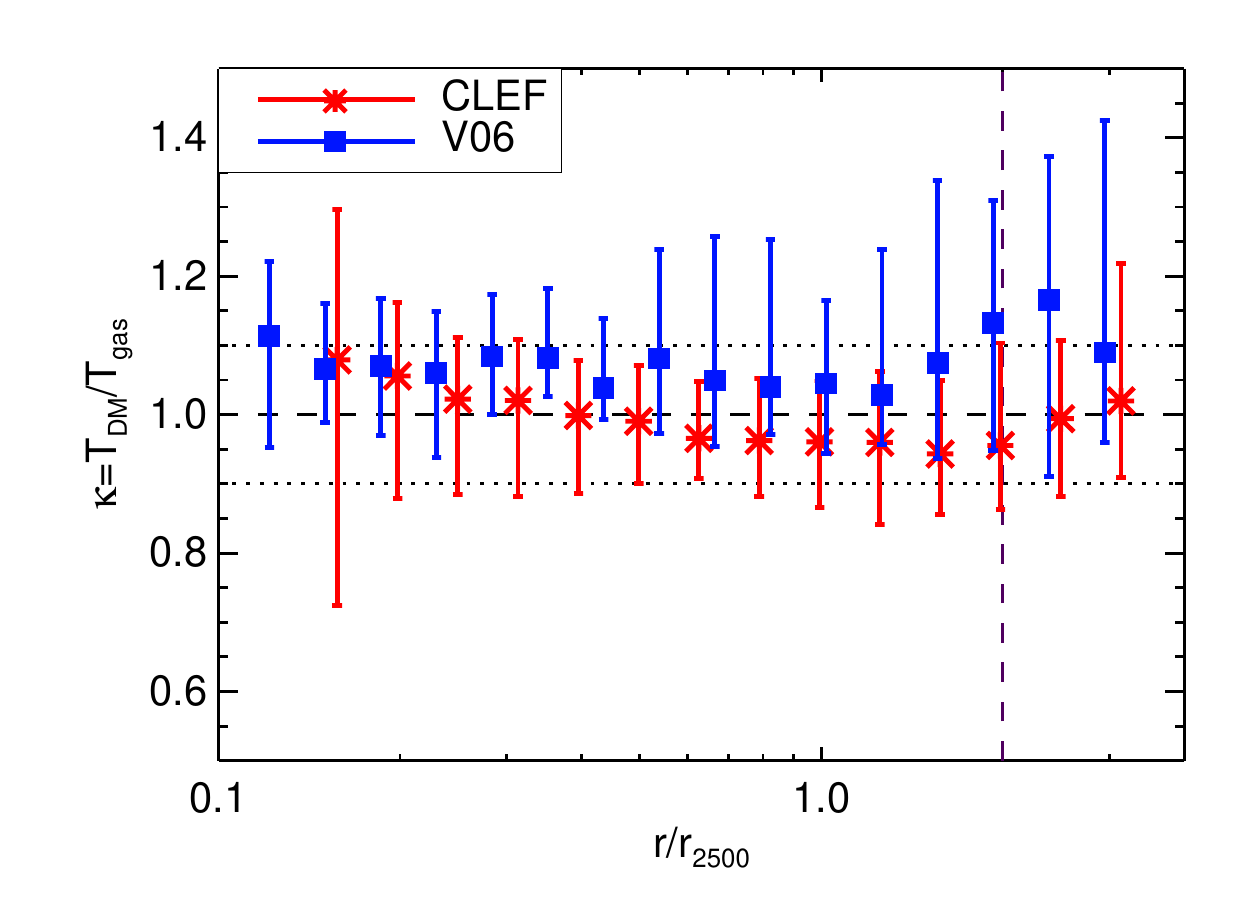}
\caption{Radial profile of $\kappa=T_{\mathrm{DM}}/T_{\mathrm{gas}}$ for the samples of clusters obtained from the CLEF and V06 simulations comprising 67 and 20 clusters, respectively. We plot the median and $1\sigma$ percentiles taken over each sample. The vertical line indicates the largest radius of the observational data sample, while the vertical lines indicate the mean and standard deviation of the $\kappa$-profile that we use in the fiducial analysis. Note that, for the CLEF sample, only eight clusters contribute to the innermost bin.}
\label{fig:kap}
\end{center}
\end{figure}

We examine the temperature relation (\ref{eq:trl}) in the two simulated samples by comparing the gas mass-weighted temperature to the rescaled dark matter velocity dispersion. The resulting $\kappa$-profiles are shown in fig.~\ref{fig:kap} and clearly $\kappa\approx1$ for both samples. Since we apply somewhat different criteria to select the two simulation samples, it is not surprising to find slightly different profiles. This indicates a systematic uncertainty of $\pm0.1$ in the simulated $\kappa$ profiles. The kinetic energy associated with bulk motions of both gas and dark matter particles is at most a few percent of the thermal energy within $2\,r_\mathrm{2500}$, outside which bulk motion is not negligible. This is in agreement with what was found in \citet{2003astro.ph..5250A}.
Due to the standard problem of limited force resolution, the simulations do not probe the innermost region reliably. Therefore we exclude data inside a cutoff radius ($56\,h^{-1}\,$kpc for CLEF, $0.1\,r_{2500}$ for V06), which means we cannot estimate $\kappa$ in the central region where gas physics can make a significant impact. 

The adiabatic version of the V06 sample exhibits a larger median $\kappa$-profile which is constant about $1.2$ within $r_{2500}$ and increases steadily to 1.4 at $r_{200}$. This is comparable with the earlier work of \citet{2004MNRAS.351..237R}, where the specific energy of dark matter was seen to be larger than that of the gas by 20--30\% in adiabatic simulations.  

\subsection{Reconstructing the velocity anisotropy}

\begin{figure}[htbp]
\epsscale{1}
\begin{center}
\plotone{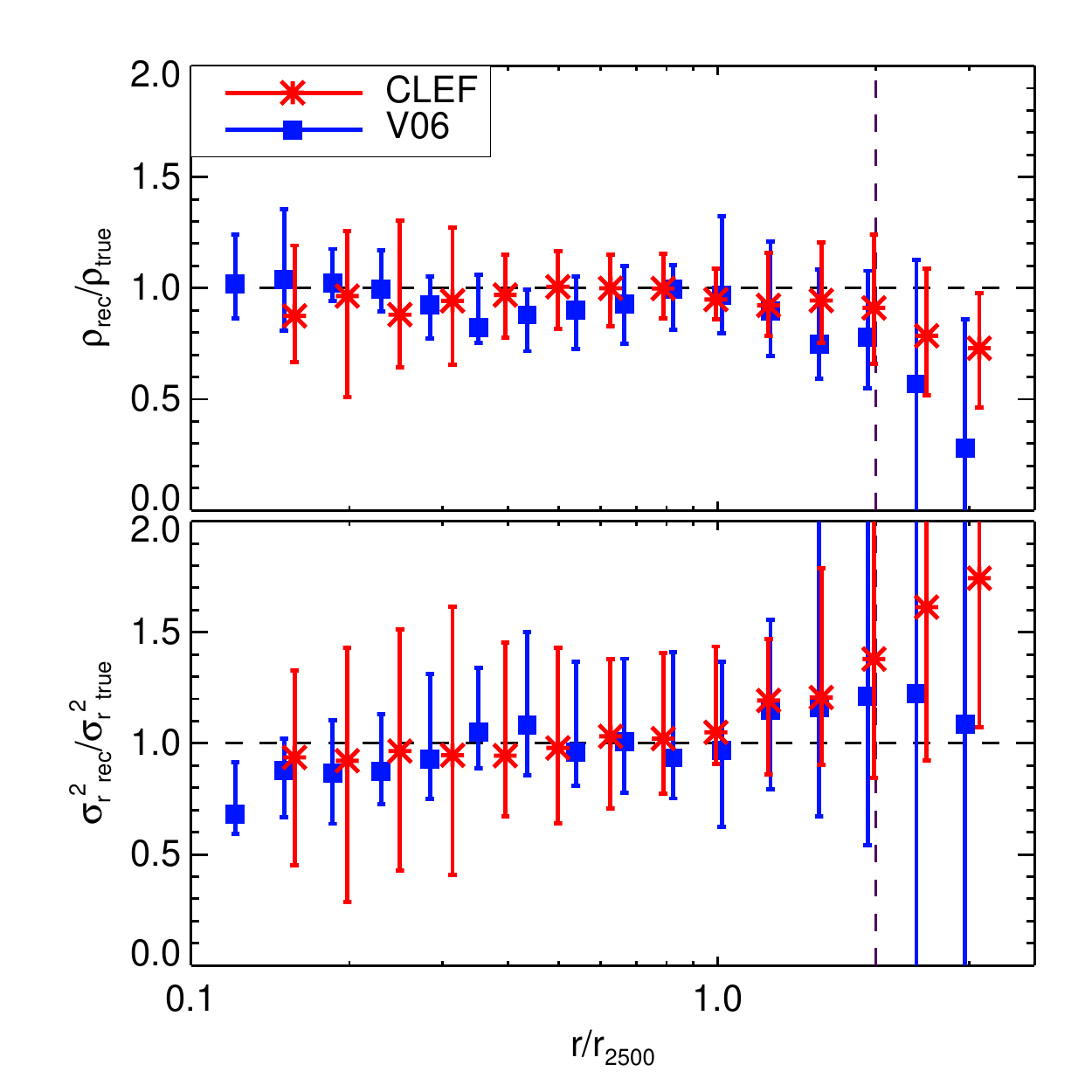}
\caption{Comparison of estimated and true values of the physical quantities involved in determining the velocity anisotropy $\beta$ in our simulations. Top, the ratio of the reconstructed total density to the true; bottom, the ratio of the reconstructed $\sigma_r^2$ to the true one. Error bars show the $1\sigma$ percentiles taken over the sample members.}
\label{fig:simsteps}
\end{center}
\end{figure}

In order to test the method outlined above for determining $\beta$, we reconstruct the anisotropy profiles observed in the simulated samples. Here, we assume $\kappa=1$ for all radii even though we expect deviations at small radii. First we derive the integrated mass profile $M(r)$ for each cluster assuming hydrostatic equilibrium (\ref{eq:he}), and from that the total density profile. The numerical derivatives involved are calculated using three-point quadratic interpolation. The estimated density profile, shown in fig.~\ref{fig:simsteps} (top), displays a satisfactory agreement with the actual density profile for both the CLEF and V06 samples. The only exception is at the outermost radii above $r_{2500}$ where the density is underestimated. Next, we calculate the radial velocity dispersion (\ref{eq:srsq}) by interpolating the integrand from $r=0$ using a four-point natural spline interpolation. We compare the resulting radial velocity dispersion with the actual in fig.~\ref{fig:simsteps} (bottom) which shows that there is good agreement except for the deviation at large radii already seen in the density profiles.

\begin{figure}[tbp]
\begin{center}
\epsscale{1}
\plotone{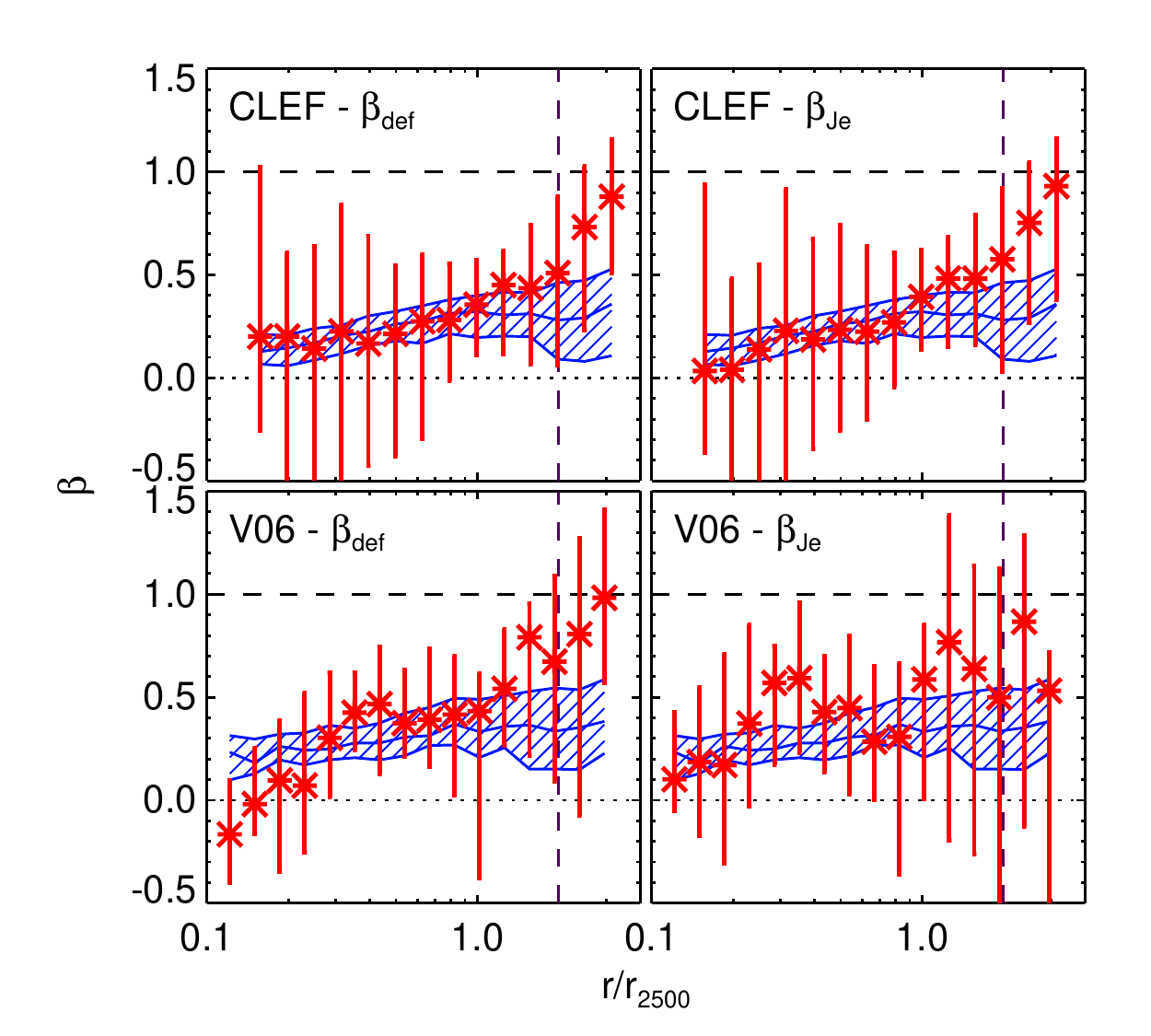}
\caption{Reconstructed velocity anisotropies for the simulated samples. The hatched bands show the actual $\beta$-profiles of the samples. Error bars show the $1\sigma$ percentiles taken over the sample members.}
\label{fig:simbeta}
\end{center}
\end{figure}

Finally we determine the velocity anisotropy parameter $\beta$. We find similar results whether we calculate $\beta_{\mathrm{Je}}$ or $\beta_{\mathrm{tr}}$, however the temperature relation yields less noisy results. The median velocity anisotropy profiles are shown in fig.~\ref{fig:simbeta} together with the median actual profile. The reconstructed profile tracks the actual anisotropy well in the inner parts but overestimates $\beta$ in the outer parts. There is also considerable noise in the results. 

\begin{figure}[tbp]
\begin{center}
\epsscale{1}
\plotone{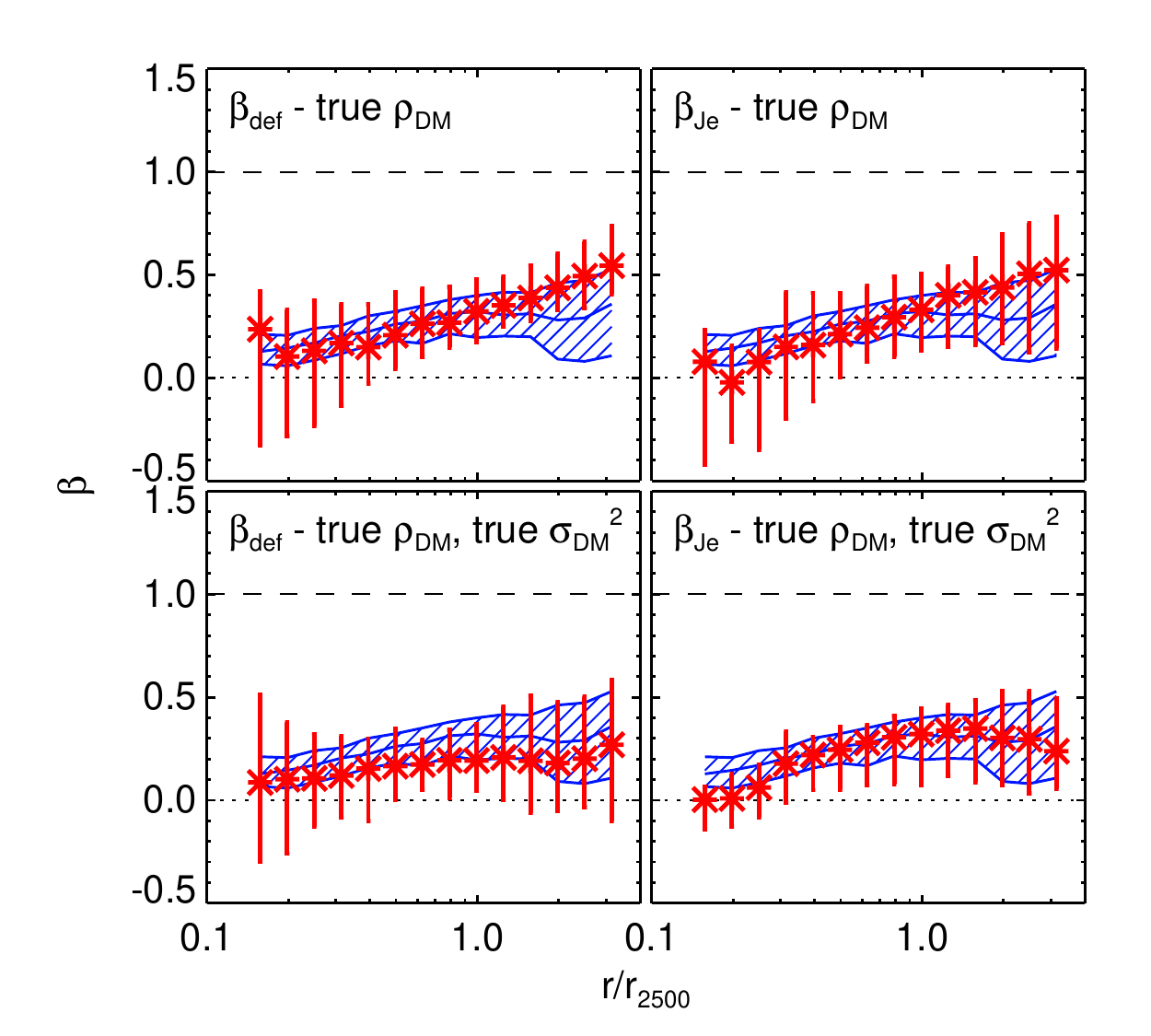}
\caption{Systematics of the reconstruction of the $\beta$ profiles for the CLEF simulation. Again, $\beta$ is recovered both from (\ref{eq:btr}) (left) and (\ref{eq:bje}) (right). The true dark matter density is substituted for the estimated, and in the bottom panels we additionally use the true total velocity dispersion instead of estimating it from $T_{\mathrm{DM}}=\kappa T_\mathrm{gas}$.}\label{fig:sys}
\end{center}
\end{figure}

In order to understand the origin of the deviations at large radii and the significant scatter in our results, we investigate the systematics of the analysis, as applied to the CLEF sample (similar conditions hold for the V06 sample). First, we substitute the dark matter density estimated from hydrostatic equilibrium with the true density. The $\beta$-profiles calculated on this basis are shown in the top panels of fig.~\ref{fig:sys}. The agreement between the estimated and actual $\beta$ is considerably improved, and the error bars are significantly reduced. This clearly indicates that, in the fiducial analysis, the numerical derivatives necessary to estimate $\rho_\mathrm{DM}$ are responsible for the large error bars. Since we do not want to do any parametrized modeling of the gas properties, the numerical derivatives are liable to amplify noise and induce systematic deviations in the outermost bin, where the quantities are only constrained to one side. Additionally, this explains why $\beta_\mathrm{Je}$ appears more noisy in the fiducial analysis since an additional derivative must be calculated. The test also shows that there is a deviation from hydrostatic equilibrium at large radii which is part of the reason why $\beta$ is overestimated. As a second test, we additionally use the true three-dimensional velocity dispersion instead of using the temperature relation. This yields further improvement as to how well the reconstructed $\beta$ tracks the true one, as shown in the bottom panels of fig.~\ref{fig:sys}. This implies that it is possible to get the correct scale of the radial velocity dispersion, calculated as an integral from the center, despite the lack of resolved data in the inner radii. We note that, with respect to observational data, the tests we apply here can possibly be utilized in the future, e.g.~with accurate density profiles inferred from gravitational lensing, and with more detailed knowledge of $\kappa$ from improved simulations. We conclude that the numerical simulations provide proof that our method is robust and that it is indeed possible to infer the $\beta$-profile despite lacking knowledge of $\kappa$ in the center.

\section{Observations}
Next we apply our analysis to observational data from which the radial gas density and temperature profiles are recovered. This is done strictly using non-parametric methods, i.e.~no modeling of the gas properties is involved. Our data consists of the deprojected density and temperature profiles of two samples of clusters at low and intermediate redshift, respectively. The deprojected profiles were obtained from X-ray data analysis published in earlier work (details below). We consider clusters which appear relaxed and close to spherical, and for which sufficient spectroscopic data are available to analyze several annuli, so that the radial variations of the gas density and temperature are resolved with good statistics. 

The first set of eleven clusters at low redshift is based on X-ray data from {\it XMM-Newton} of the clusters: A262, A496, A1795, A1837, A2052, A4059, S\'ersic 159$-$3, MKW3s, MKW9, NGC533, and 2A0335+096. These objects are highly relaxed cool-core (CC) clusters selected as to match the requirements described above. The objects were part of the sample analyzed in \citet{2004A&A...413..415K} (see this paper for an extensive presentation of the data analysis), in which deprojected radial temperature and density profiles were derived from spatially resolved spectroscopy. We adopt the radial bin selection of \citet{2005A&A...433..101P} in order to ensure a robust determination of gas temperature and density for the full radial range. Note that data for A2052 and S\'ersic 159$-$3 were also used in the analysis by \citet{2007A&A...476L..37H} where a constant velocity anisotropy was assumed.

The other set of five intermediate redshift X-ray galaxy clusters (RXJ1347.5, A1689, A2218, A1914, A611) is from the {\it Chandra} sample analyzed in \citet{Morandi:2007aw}. The radial deprojected temperature and density profiles were retrieved through resolved spectral analysis in a set of annuli, selected to collect at least 2000 net counts, by assuming spherical geometry and by using the definition of `effective volume' (see \citet{Morandi:2007aw} for further details).

\section{Results}\label{sec:res}

\begin{figure*}[tbp]
\begin{center}
\epsscale{1}
\plotone{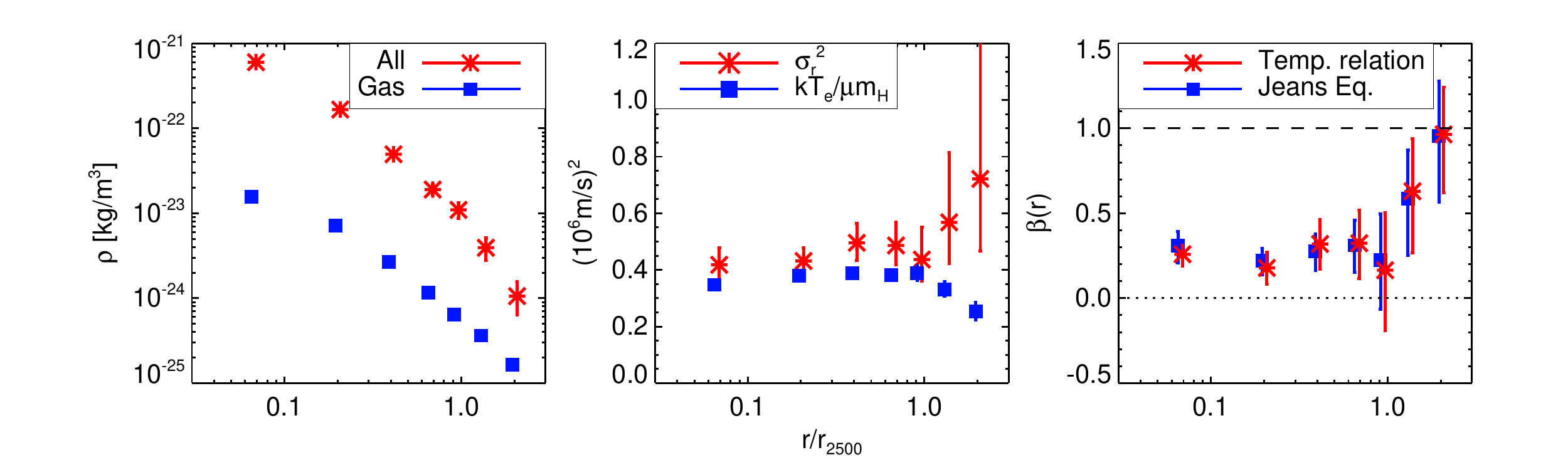}
\caption{Three steps in the calculation of the velocity anisotropy for S\'ersic~159$-$3. Left, the inferred total density; center, the radial velocity dispersion; right, the $\beta$-profiles. The gas density and temperature profiles are also shown. The scale radius for this cluster is estimated to be $r_{2500}=337\pm13\,$kpc. Error bars indicate the propagated statistical uncertainties on the ICM temperature and density profile, taken as the $1\sigma$ percentiles of 1000 Monte Carlo samples. This is unlike in the previous figures where the error bars indicate the spread over the numerically simulated samples. In the right panel, the radial positions of $\beta_\mathrm{tr}$ and $\beta_\mathrm{Je}$ have been offset slightly for clarity.}
\label{fig:beAS}
\end{center}
\end{figure*}

\begin{figure*}[htbp]
\begin{center}
\epsscale{1}
\plotone{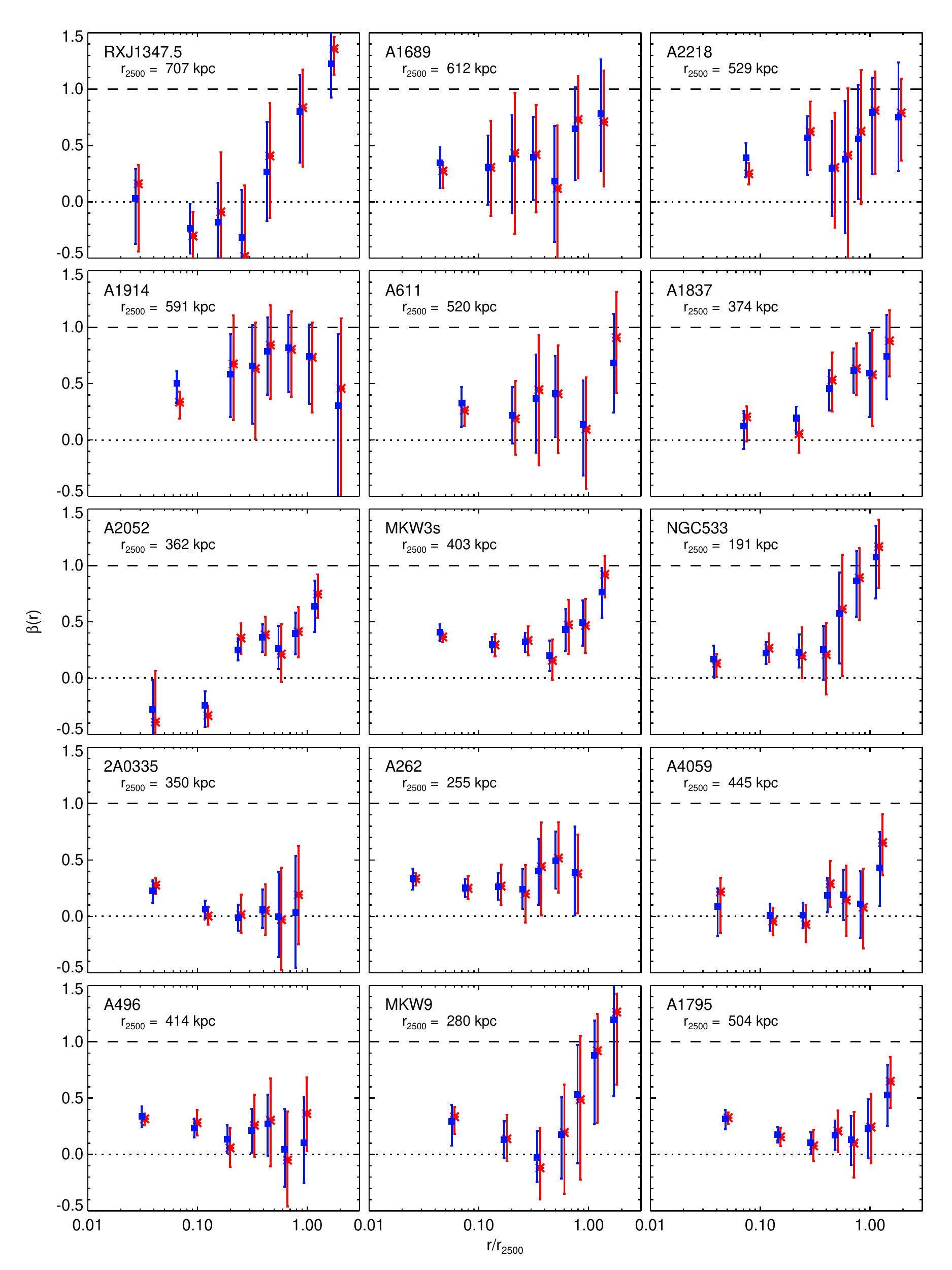}
\caption{Median velocity anisotropy profiles for the remaining 15 clusters of our sample. The estimated scale radii are also shown, and the symbols are the same as in fig.~\ref{fig:beAS}.}
\label{fi:bsmpl}
\end{center}
\end{figure*}

We determine the dark matter velocity anisotropy profile $\beta(r)$ of each cluster according to the recipe in section \ref{se:th} using a Monte Carlo method. For each radial bin the deprojected gas temperature and density are sampled assuming Gaussian uncertainties, i.e.~a random number is chosen from a Gaussian distribution with mean equal to the estimated temperature or density and a standard deviation equal to the uncertainty of the estimate. The bins are sampled independently. The parameter $\kappa$ is also sampled for each bin, assuming a Gaussian distribution with a mean of 1 and a standard deviation of 0.1, which is a reasonable value according to the simulations. The sampled profiles are used to reconstruct the total mass through (\ref{eq:he}), and then the integrand, the radial velocity dispersion, and the velocity anisotropy are calculated in each bin. The sampled set of profiles is accepted only if the temperature and density as well as the reconstructed dark matter density and radial velocity dispersion are all non-negative in all bins. For each sample, we also estimate the scale radius $r_{2500}$ and the mass $M_{2500}$ contained within that radius. Table \ref{tb:prop} summarizes the properties of the clusters in our sample.

The numerical methods for calculating derivatives and integrals are the same as for the simulated samples, i.e.~three-point quadratic interpolation is used for derivatives and four-point spline interpolation is used for the integral in (\ref{eq:srsq}). The integration results are stable to using two-point linear, three-point quadratic, or four-point least squares quadratic interpolation instead. 

Individual steps of the reconstruction are shown in fig.~\ref{fig:beAS} for the cluster S\'ersic~159$-$3, and the deprojected input data are also displayed. We always plot the median and $1\sigma$ percentiles since spurious outliers in individual Monte Carlo samples can bias the mean and standard deviation significantly. The size of the error bars is mostly determined by the uncertainties of the temperatures, to a lesser degree by the uncertainties of the ICM densities, and it is virtually insensitive to the 10\% variation assumed for the $\kappa$-profile.

As can be seen in the right panel of fig.~\ref{fig:beAS}, the agreement between $\beta_\mathrm{tr}$ and $\beta_\mathrm{Je}$ indicates that numerical effects associated with the integration and differentiations are small. On the other hand, $\beta$ becomes unphysically large in the outermost bins since the reconstructed radial velocity dispersion for some samples becomes greater than the total velocity dispersion. This result is similar to that found in the blind analysis of the simulation samples. As discussed above, this behaviour is mainly due to a deviation from hydrostatic equilibrium of the gas, and to a lesser degree to edge effects making the numerical differentiations less well determined in the outermost bin. It is possible that systematic uncertainties in the input data or radial variations in $\kappa$ for individual clusters also play a role. In principle, we could impose $\sigma_r^2<\sigma^2$, thereby forcing $\beta<1$, as another physical condition on each Monte Carlo sample, but we prefer not to do so in order to have a consistency check.

We repeat the data analysis for the remaining 15 clusters of our sample and the resulting velocity anisotropy profiles are shown in fig.~\ref{fi:bsmpl}. In almost all cases the anisotropy is small in the inner radial bins and increases to between $0.5$ and $1.0$ in the outer parts. There is good agreement between the two derivations of $\beta$ for all clusters, indicating that numerical issues are under control.

\begin{figure}[htbp]
\begin{center}
\epsscale{1}
\plotone{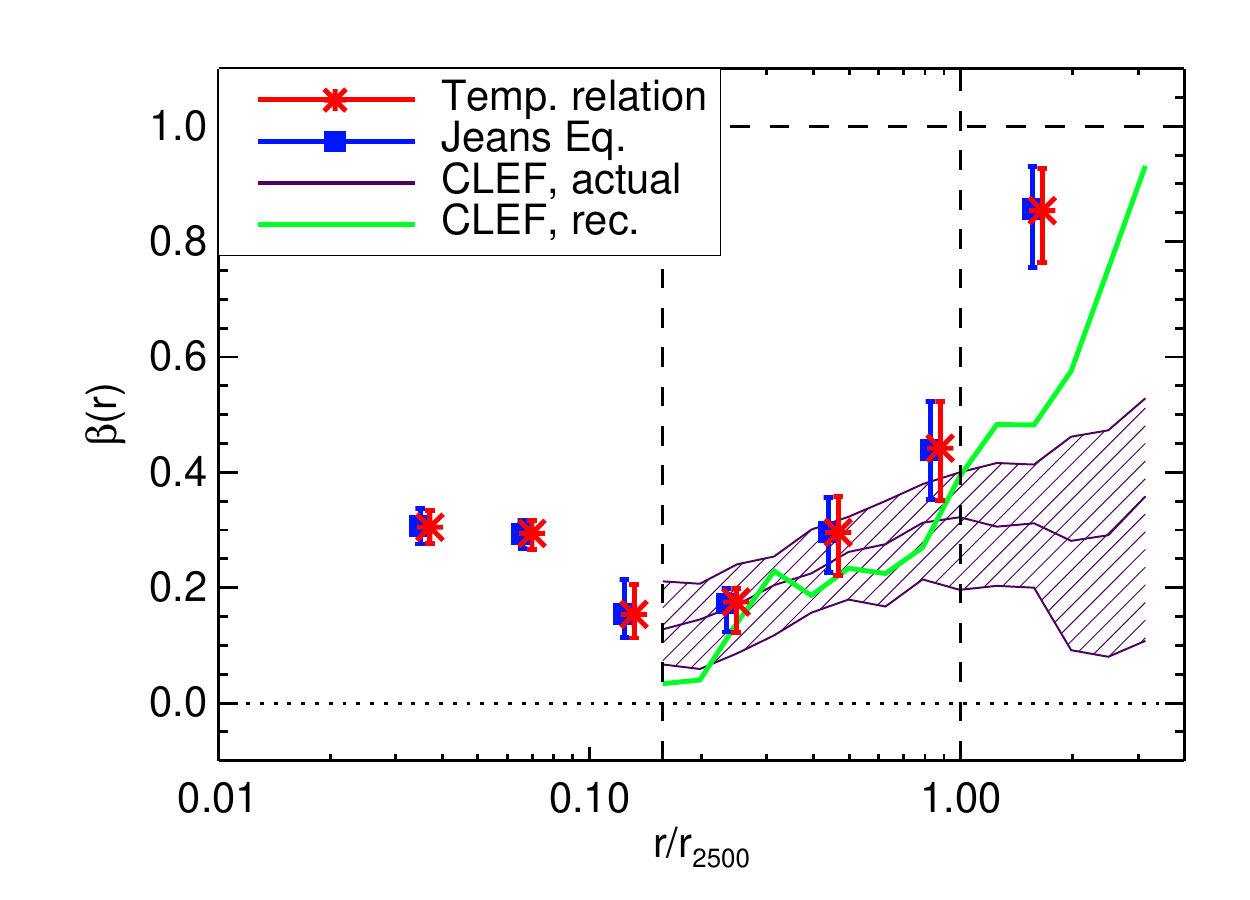}
\caption{Median velocity anisotropy profile of all 16 clusters in our dataset. In this case the error bars denote the $1\sigma$ percentiles of the combined probability density of all clusters within the bin. The actual and reconstructed $\beta$-profiles from the simulations are also shown. The left vertical line is the innermost radius probed in the CLEF simulations and the right vertical line shows, roughly, the onset of significant deviations from hydrostatic equilibrium in the simulations, see fig.~\ref{fig:simsteps}.}
\label{fi:stack}
\end{center}
\end{figure}

Since the qualitative behaviour of the velocity anisotropy profiles are similar, we combine all our data into a single `stacked' profile, shown in fig.~\ref{fi:stack}. In the region where direct comparison is possible, the measured stacked  profile is very similar to the reconstructed $\beta$ profiles for the simulation samples (the green line), and within $r_{2500}$ there is also agreement with the actual velocity anisotropy of the simulation samples (hatched band). The velocity anisotropy is likely overestimated outside $r_{2500}$ for the same reason as for the simulated samples, i.e.~deviation from hydrostatic equilibrium, but the effect appears to be even stronger for the observational data. Interior to the cut-off radius of the numerical simulations, the observations tend to $\beta\sim0.3$. This is somewhat surprising since numerical simulations at all mass scales generally have very little anisotropy towards the center of structures. While we cannot exclude the possibility that cluster halos are anisotropic even at low radii, our result can also be explained by the neglected stellar contribution $\rho_\star$ to the total mass density. To first order, this contribution enters our analysis in the Jeans equation through the estimated dark matter density $\widetilde{\rho}_\mathrm{DM}=\rho_\mathrm{DM}+\rho_\star$. In terms of $\delta_\star=\rho_\star/\rho_\mathrm{DM}$, the Jeans equation becomes
\begin{eqnarray}
\sigma_r^2\left(\frac{d\ln \widetilde{\rho}_{\mathrm{DM}}}{d\ln r}+\frac{d\ln \sigma_r^2}{d\ln r}+2\beta-\frac{d\ln (1+\delta_\star)}{d\ln r}\right)&&\nonumber\\=-\frac{GM(r)}{r}&&,
\end{eqnarray}
where the slope of $(1+\delta_\star)$ is negative since the stellar density must fall off faster than the dark matter density. This means that we overestimate the velocity anisotropy in the central region by not accounting for the stellar mass. Indeed, if we assume that 50\% of the total mass in the innermost bin is made up of stars, the velocity anisotropy in the two innermost bins becomes consistent with zero. There is also a second order correction through the appearance of $\widetilde{\rho}_\mathrm{DM}$ in (\ref{eq:srsq}) instead of $\rho_\mathrm{DM}$, but this correction must be small since the density contributes to both the integrand and the normalization factor.

\begin{deluxetable}{rccc}
\tablewidth{0pt}
\tablecaption{Properties of our cluster sample\label{tb:prop}}
\tablehead{\colhead{Cluster}&{$z$}&{$r_{2500}/$kpc}&{$M_{2500}/M_\odot$}}
\startdata
A262 & 0.015 & $256\pm28$ & $(2.7\pm0.8)\times10^{13}$\\
A496 & 0.032 & $398\pm10$ & $(1.0\pm0.2)\times10^{14}$\\
A1795 & 0.064 & $504\pm22$ & $(1.9\pm0.2)\times10^{14}$\\
A1837 & 0.071 & $374\pm26$ & $(8.0\pm1.7)\times10^{13}$\\
A2052 & 0.036 & $362\pm11$ & $(6.7\pm0.6)\times10^{13}$\\
A4059 & 0.047 & $445\pm21$ & $(1.3\pm0.2)\times10^{14}$\\
S\'ersic~159$-$3 & 0.057 & $337\pm17$ & $(5.7\pm0.8)\times10^{13}$\\
MKW3s & 0.046 & $404\pm14$ & $(9.5\pm0.9)\times10^{13}$\\
MKW9 & 0.040 & $279\pm44$ & $(3.2\pm1.5)\times10^{13}$\\
NGC533 & 0.018 & $191\pm15$ & $(9.7\pm2.2)\times10^{12}$\\
2A0335+096 & 0.034 & $350\pm40$ & $(6.9\pm2.5)\times10^{13}$\\ \hline
A611 & 0.29 & $519\pm52$ & $(2.5\pm0.6)\times10^{14}$\\
A1689 & 0.18 & $609\pm4$ & $(3.5\pm0.7)\times10^{14}$\\
A1914 & 0.17 & $590\pm44$ & $(3.3\pm0.8)\times10^{14}$\\
A2218 & 0.18 & $535\pm51$ & $(2.5\pm0.7)\times10^{14}$\\
RXJ1347.5-1145 & 0.45 & $710\pm60$ & $(7.3\pm1.4)\times10^{14}$
\enddata
\end{deluxetable}

\begin{figure*}[tbp]
\begin{center}
\epsscale{1}
\plotone{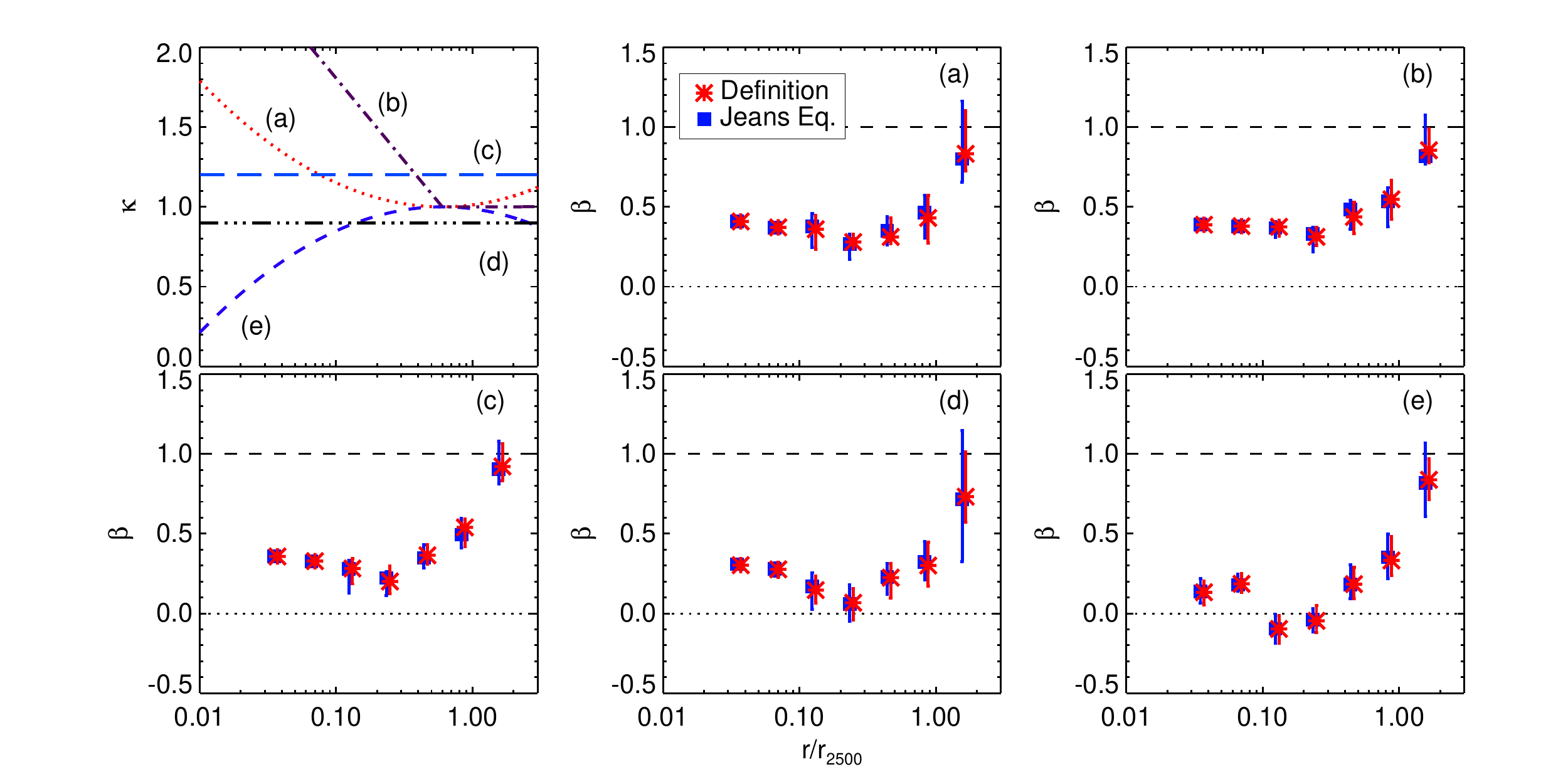}
\caption{The effect of assuming different $\kappa$-profiles on the stacked velocity anisotropy profile. Top left, the five $\kappa$-profiles. Others, the resulting sample averaged $\beta$-profiles calculated assuming the numbered $\kappa$-profile. In all cases, $\beta$ is greater than zero in the outer parts. }
\label{fi:ksys}
\end{center}
\end{figure*}

Finally, we investigate how the assumed shape of the $\kappa$-profile affects our results. We try five different profiles as functions of $x=r/r_{2500}$ with noise added as before, and calculate the velocity anisotropy profiles for each. The $\kappa$-profiles are chosen so as to mimick either the effects of gas radiative cooling or AGN heating in the central regions, or to check the results if the dark matter is generally hotter or cooler than the gas. The radially varying profiles we try are extreme cases of the simulation profiles, fig.~\ref{fig:kap}. Typically, the result is that the $\beta$-profile is shifted in the central regions while the outer regions are largely unaffected, as shown in fig.~\ref{fi:ksys}. This analysis confirms that there is a significant velocity anisotropy at large radii, independent of the specific assumptions about the  temperature relation.

\section{Summary and discussion}
In this paper, we have presented a non-parametric method to infer the velocity anisotropy of dark matter in clusters of galxies from the observable temperature and density of the intracluster medium. We assume that the intracluster medium has the same specific energy as the dark matter, and we investigate the validity of this assumption in two different cosmological simulations of the formation of galaxy clusters. Both confirm the simplest possible form of the relation, namely $T_\mathrm{DM}\approx T_\mathrm{gas}$ in the radial range which is resolved. 

We have tested how well our method can reconstruct the actual velocity anisotropy in the simulated clusters, and we have found good agreement between the two, although the reconstruction is sensitive to systematic biases connected with deviations from hydrostatic equilibrium. 

We have applied our method to the radial ICM density and temperature profiles of 16 galaxy clusters based on {\it Chandra} and {\it XMM-Newton} X-ray data. The shape of the velocity anisotropy profiles is always consistent with that seen in simulations, which tends to zero at the innermost radius where the temperature relation is calibrated. It then increases to about 0.5 at $r_{2500}$ and even larger in the outer regions. The same is true of the fiducial analysis applied to simulated data and is likely caused by a deviation from hydrostatic equilibrium outside $r_{2500}$. We also find a significant anisotropy even if we assume radially varying $\kappa$-profiles, such as can be expected given the strong gas cooling and AGN heating in the core of many clusters, or if we assume $\kappa\ne 1$. The agreement between the observed velocity anisotropy and that predicted in numerical simulations shows that we are beginning to understand also the dynamical aspects of dark matter in halos.

In the innermost radial bins we measure a rather large anisotropy, but this is most likely an overestimation due to the neglect of the stellar mass in the center. This can be used as a means to estimate the stellar mass profile of galaxy clusters if one assumes that the velocity dispersion to be isotropic in the central regions. Similarly, our method may be used as a general test of whether a cluster is relaxed. A reconstructed velocity anisotropy which deviates significantly from the simulated profiles would be a strong hint that the data do not support the assumption of hydrostatic equilibrium.

The inferred velocity anisotropy profiles are significantly different from zero which means that the collective behaviour of dark matter is unlike that of baryonic particles in gases. This shows that dark matter is effectively collisionless on the timescale of $\tau\sim10^9$, the dynamical timescale of galaxy clusters. By taking typical values at $\sim0.3\,r_{2500}$ and allowing only a few scatterings within the time $\tau$, this corresponds to an order--of--magnitude upper limit to the scattering cross-section of roughly $\sigma/m=(\rho_{\textrm{DM}}\tau v)^{-1}\lesssim1\,$cm$^2$g$^{-1}$. This limit is similar to what has been found for merging clusters \citep{2004ApJ...606..819M,2008arXiv0806.2320B}, and within an order of magnitude of the scattering cross-section for self-interacting dark matter proposed in \citet{Spergel:1999mh}.

We emphasize that improvements to the numerical simulations in the near future will improve our understanding of the $\kappa$ profile and hopefully track the impact of radiative effects in the center. We also hope that improved understanding of deviations from hydrostatic equilibrium will allow us to estimate how large the suspected bias at large radii is. On the observational side, the main problem at present is the uncertainty in the temperature profile. Improvements can be expected both with regards to the deprojection analysis and the amount of data available. Obviously, there is also the possibility of including a kinematical analysis of the galaxy clusters in our method.

\acknowledgments
We thank Jens Hjorth, Gary A.~Mamon, and Kristian Pedersen for comments. The Dark Cosmology Centre is funded by the Danish National Research Foundation. SE acknowledges the financial contribution from contract ASI-INAF I/023/05/0 and I/088/06/0.

{\it Facilities:} \facility{XMM}, \facility{CXO}

\bibliographystyle{apj}
\bibliography{betacluster}
\end{document}